\documentclass[aipapl,numerical,reprint]{revtex4-1}
\usepackage{graphicx}

\begin{document}


\title{Standoff Detection via Single-Beam Spectral Notch Filtered Pulses} 


%
\author{Adi Natan, Jonathan M. Levitt, Leigh Graham, Ori Katz, and Yaron Silberberg}
%
\affiliation{Department of Complex Systems, Weizmann Institute of Science, Rehovot 76100, Israel}


\date{\today}

\begin{abstract}
We demonstrate single-beam coherent anti-Stokes Raman spectroscopy (CARS), for detecting and identifying traces of solids, including minute amounts of explosives, from a standoff distance ($>$50 m) using intense femtosecond pulses. Until now, single-beam CARS methods relied on pulse-shapers in order to obtain vibrational spectra. Here we present a simple and easy-to-implement detection scheme, using a commercially available notch filter, that does not require the use of a pulse-shaper.
\end{abstract}


\maketitle 


Laser based remote detection and identification of hazardous materials, including biological warfare agents and explosives has recently been the focus of intense research efforts \cite{katz2008ref1,Wallin2009,dantus2011apl,scully_back_lasing,Carter2005,Bar2008,Pestov2008}. In particular, it has been shown that the vibrational response of molecules provides a unique fingerprint for various molecular species and, hence, vibrationl spectra can be used for identification \cite{Carter2005,Bar2008,Pestov2008,Dantus2008,Scrader,Pestov2007,katz2008}. One such suitable technique for indenting vibrational spectra is coherent anti-Stokes Raman spectroscopy (CARS) \cite{Volkmer2005}. In CARS, pump $\omega_{p}$ and Stokes $\omega_{s}$ photons excite a vibrational level $\omega_{vib}$ that is subsequently probed with a probe photon $\omega_{pr}$. The vibrational levels are resolved by measuring the scattered anti-Stokes photons located at $\Omega_{vib}+\omega_{pr}=\omega_{AS}$. In conventional CARS, narrowband (ps) pump and probe beams are tuned such that $\omega_{p}-\omega_{s}=\Omega_{vib}$ where a \emph{single} vibrational level is excited. Multiplex CARS can also be achieved by using spectrally broader (fs) Stokes pulses to excite \emph{multiple} bands of vibrational levels and identify them by measuring the spectrally resolved anti-Stokes photons.

CARS is typically challenging to implement due to the strict spatiotemporal overlap requirement of several beams from several sources \cite{Scrader,Xie2008,Volkmer2005}. To overcome this issue, single-beam techniques have been developed using shaped, spectrally broad femtosecond pulses to simultaneously provide the necessary pump, Stokes, and probe photons \cite{Dudovich2002Nature,OronPhaseContrast,Oron2003phase_and_pol,Dudovich2003,Leone2007,Oron2002,Motzkus2009}. By employing single-beam techniques, spatiotemporal overlap is inherently achieved; however, the exclusive use of temporally short, femtosecond pulses results in a nonresonant four-wave mixing (FWM) signal that is orders of magnitude greater than the resonant CARS signal \cite{OronPhaseContrast,Leone2007}. The measured CARS signal is therefore the coherent sum of the resonant and nonresonant signals (Equation (1)).

A variety of CARS schemes have been developed in order to suppress the nonresonant background
signal, such as tailoring the temporal width and
delay of the probe-pulse \cite{Pestov2007,Oron2002}, using polarization pulse shaping techniques \cite{Oron2003phase_and_pol}, and employing coherent control \cite{Oron2002,Dudovich2002Nature}. These schemes exploit different properties of the nonlinear FWM interaction to achieve the goal of background-free measurements.
While these methods are effective for applications such as microscopy \cite{Dudovich2002Nature,Xie2006}, they are not necessarily suitable for long-range standoff probing of scattering samples. Under such conditions, polarization shaping is sensitive to the depolarization caused by the multiple random scattering in the sample. Furthermore, the nonresonant signal can serve to amplify the resonant signal and, thus, suppression can result in a remaining resonant signal that is too weak to be detected.

In this work, we demonstrate single beam standoff detection of traces of solids without using a complex pulse shaping apparatus or nonresonant background suppression. Instead, we use a commercially available resonant photonic crystal slab filter (RPCS) \cite{afm} as passive optical element and exploit the nonresonant signal in a homodyne amplification scheme.

The RPCS is a waveguide etched with a subwavelength grating, that creates a tunable notch feature in the spectrum (Fig. 1(a-b)). When a broadband light impinges the RPCS at a given angle and polarization, most wavelengths are fully transmitted. However, a narrow resonant band is diffracted by the grating and coupled to the waveguide layer. The coupled light interferes destructively with the transmitted wave, resulting in the reflection of the resonant band and the creation of a notch feature in the transmission spectrum. In the time domain, the spectral notch is translated to a delayed narrowband pulse. The notch frequency, i.e. spectral band coupled into the waveguide, can be adjusted over a wide spectral range by changing the angle of the RPCS with respect to the incident beam.

\begin{figure}
\begin{center}
\includegraphics[width=\columnwidth]{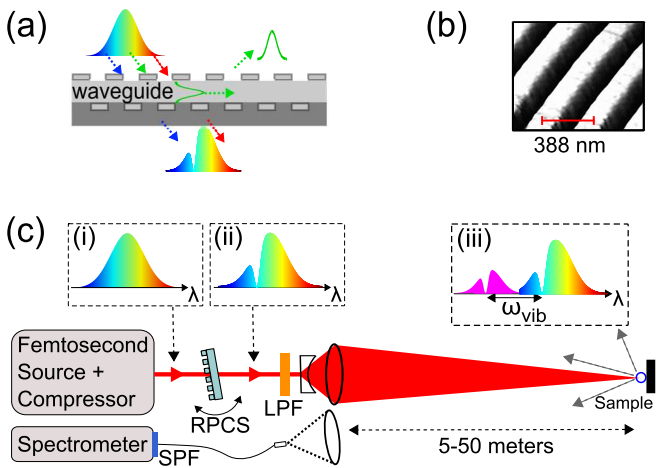}
\end{center}
\caption{(color online) (a) Schematic diagram of the RPCS filter, see text for details. (b) Atomic force microscopy
image of the RPCS surface (from \cite{afm}). (c) The experimental setup. The broadband excitation pulse (i), is shaped with a tunable narrowband spectral notch by the RPCS filter (ii). A long-pass filter (LPF) is then used to attenuate the short-wavelength region. The spectral notch serves as a probe for the CARS process, generating narrow features which are blue-shifted from the probe by the vibrational frequencies (iii). The scattered light is collected with a lens, then short-pass filtered (SPF) and coupled to a spectrometer.} \label{fig_1}
\end{figure}

The use of an RPCS as a passive shaping element has been recently demonstrated in single-pulse CARS mircospectroscopy \cite{notch}.
In this technique, the notched-shaped spectral feature serves as a narrow probe that allows for multiplex CARS detection \cite{notch,Natan2009}.  We exploit the strong nonreasonant background and use it as a local oscillator in a homodyne amplification scheme:
\begin{eqnarray}
    I(\omega) &=& |P^{NR}(\omega)+P^{R}(\omega) |^2  \\
   \nonumber &\simeq& |P^{NR}(\omega)|^2+2 |P^R(\omega)||P^{NR}(\omega)| \cos{\phi(\omega)}
\end{eqnarray}
where $I(\omega)$ is the measured signal intensity at frequency $\omega$, $P_R$
and $P_{NR}$ are the resonant and nonresonant signals, respectively, and $\phi$ is the relative spectral phase between them. Usually, with ultrashort pulses $|P^{NR}(\omega)|^2 \gg |P^{R}(\omega)|^2$ , making the the resonant signal challenging to measure \cite{Pestov2007,OronPhaseContrast,Leone2007}. However, the cross term is the product of a strong, smooth and featureless nonresonant signal, with a weak resonant signal that has a typical spectral phase modulation around a vibrational resonance. The phase modulation is governed by the vibrational resonance lineshape and the phase structure of the RPCS notch resonance \cite{Oron2002}. As a result, the resonant feature is amplified by several orders of magnitude, allowing for detection of the much weaker coherent Raman signal.

\begin{figure}
\begin{center}
\includegraphics[width=\columnwidth]{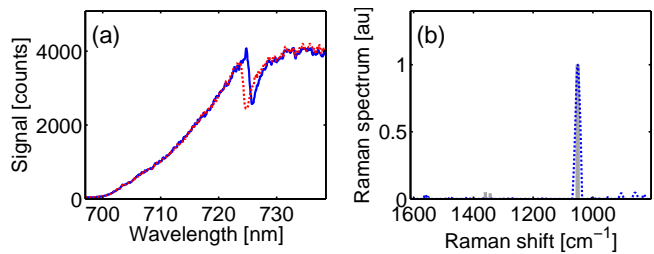}
\end{center}
\caption{(color online) (a) Raw spectra of trace amount ($<$1 mg) of KNO$_3$ collected from a distance of 5 meters, using two notch positions: 784 nm (dashed red) and 785 nm (solid blue) with integration time of 3 seconds. (b) Taking the normalized difference of the two spectra by the non-resonant spectra (dashed blue), the Raman line (solid gray) is retrieved.} \label{fig_2}
\end{figure}

To demonstrate the ability to remotely detected backscattered CARS signal, we probed samples with 30 fs pulses (Femtolasers GmbH) that were notch filtered by an RPCS. The pulses were then long-pass filtered around the spectral region that overlapped with the CARS signal, and focused on samples at distances of 5-50 meters (Fig. 1(c)). We placed a black, absorptive material behind the samples to eliminate the possibility of collecting reflections of the much stronger forward propagating CARS signal. This is in contrast to previous studies that placed the sample on a reflective surface \cite{Dantus2008}. The beam diameter on the sample was $\simeq$ 1.5 mm, resulting in peak powers of $\simeq 10^{11} \text{W/cm}^2$. We collected the backscattered CARS signal from various materials using a single 7.5$''$ lens imaging configuration. The light was then short-pass filtered and fiber coupled to a spectrometer (Horiba Jobim Ivon, Triax 321). For each acquisition we obtained two measurements, corresponding to different notch positions, spectrally separated by the notch width ($\simeq$ 1 nm). The CARS spectra were obtained with integration times ranging from 1-5 seconds. The dispersion induced by the 5-50 m propagation in air was compensated using the internal prism compressor of the laser source.

Typical CARS spectra from complimentary notch locations are shown in Fig \ref{fig_2}. The interference feature due to the resonant contribution at 1051 cm$^{-1}$ is evident in the KNO$_3$ spectrum. In order to extract the Raman line, we normalized the differential measurement by P$^{NR}$ \cite{notch}:
\begin{equation}
I_{CARS}(\omega) \propto \frac{I_{n(\lambda+\Delta \lambda )}(\omega)} {P_{n(\lambda+\Delta \lambda )}^{NR}(\omega)} -\frac{I_{n(\lambda)}(\omega)}{P_{n(\lambda)}^{NR}(\omega)}
\end{equation}
Where I$_{n(\lambda)}$ is the intensity measured with a notch at wavelength $\lambda$, of width $\Delta \lambda$, and P$^{NR}$ is approximated by fitting a smooth curve to $\sqrt{I_{n(\lambda)}(\omega)}$.

\begin{figure}
\begin{center}
\includegraphics[width=1\columnwidth]{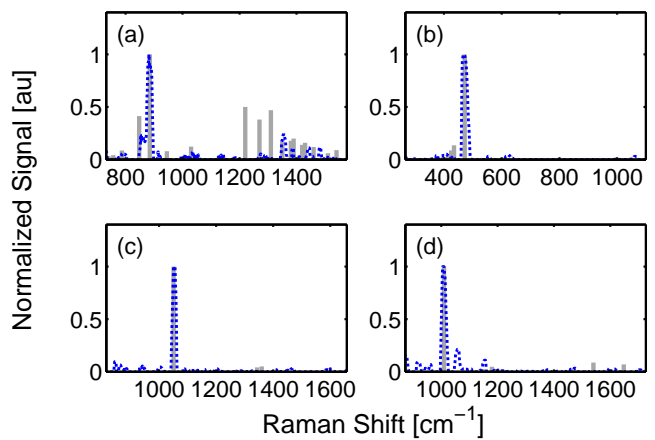}
\end{center}
\caption{Resolved CARS spectra of several scattering samples (dashed blue). (a) Cyclotrimethylene-trinitramine (RDX/T4) $<$1 mg (b) Sulfur powder $<$500 $\mu$g (c) Crystalized Potassium Nitrate (KNO$_3$) $<$1mg (d) Crystalized Urea $<$ 4 mg.  Spectra were obtained at a standoff distances of (a) 24 m and (b-d) 50 m, with integration times of (a) 3 s (b) 5 s (c) 5 s (d) 1 s. The Raman lines (solid gray) are plotted for reference.} \label{fig_3}
\end{figure}

Experimentally resolved Raman spectra of traces of solids and explosives particles are presented
in Fig. \ref{fig_3}. The measured spectra are in good agreement with the known Raman lines
of the probed materials \cite{katz2008}. In our experiments, the acquisition time was limited by the the beam-point stability of the experimental setup. Actively stabilizing the beam should improve SNR due to variations between measurements, and allow shorter measurement times.
Many alternative shaping techniques could also be used to generate a similar spectral notch-like feature. For example, the use of an interference filter, Fabry-Perot interferometer, or fiber Bragg grating would yield similar results. However, these lack the simplicity, compactness, and compatibility with high intensities.

Femtosecond CARS spectroscopy using a spectral notch filter has numerous advantages over other single beam techniques. The RPCS is virtually alignment free and has broad tunability allowing for flexibility in the detection window, while nearly eliminating the inherent losses of grating based pulse shapers. A transmission of almost unity is particularly beneficial as the CARS signals are cubically proportional to the pulse intensity. Moreover, the RPCS can be mounted on a galvanometric mirror to achieve rapid modulation rates that are one to four orders of magnitudes faster than typical LC-SLMs, allowing lock-in detection and fast scanning. Additional signal enhancement can be made using multiple notch filters. By spectral positioning of several notches spaced by the vibrational lines of a known substance, the coherent addition of signals from several vibrational levels will generate a feature which is significantly larger than the linear sum of the individual contributions \cite{Oron2004}.

We thank S. Soria for providing the RPCS filter and the AFM image used in this work \cite{afm}. We also thank E. Grinvald, G. Elazar and G. Han for invaluable help. This work was supported in parts by DHS Center of Excellence of Explosives Detection, Mitigation, Response and by NATO project no. SfP-983789. JML was supported by the European MC-ITN FASTQUAST.

\end{document}